\documentclass[mathleft
]{an}
\usepackage{graphicx}
\usepackage{times}
\begin{document}

% The following seven commands are intended for editorial usage and should be ignored by
% the author(s).
\Pagespan{789}{}% Document's page range. 
% If second parameter is left empty, the last page is computed automatically.
\Yearpublication{2010}%
\Yearsubmission{2010}%
\Month{}%   
\Volume{999}%  
\Issue{88}% 
% \DOI{This.is/not.aDOI}% 

\title{Can an underestimation of opacity explain B-type pulsators in the SMC?}

\author{S. Salmon\inst{1,2}\fnmsep\thanks{Corresponding author:
  \email{sebastien.salmon@doct.ulg.ac.be}\newline}
%Example 
%for footnote, note the usage of the \texttt{fnmsep}
%command as separator between institute number and footnote mark} 
\and J. Montalb\'{a}n \inst{1}
\and A. Miglio \inst{1,3}
\and M-A. Dupret \inst{1}
\and T. Morel \inst{1}
\and A. Noels \inst{1}
}
\titlerunning{Instructions for authors}
\authorrunning{T.H.E. Editor \& G.H. Ostwriter}
\institute{
Institut d'Astrophysique et de G\'{e}ophysique de l'Universit\'{e} de Li\`{e}ge, All\'{e}e du 6 Ao\^{u}t 17, B-4000 Li\`{e}ge, Belgium
\and
Fonds pour la formation \`{a} la Recherche dans l'Industrie et dans l'Agriculture-FRIA, Belgium
\and 
Postdoctoral Researcher, Fonds de la Recherche Scientifique-FNRS, Belgium
}

\received{30 May 2005}
\accepted{11 Nov 2005}
\publonline{later}


\abstract{Slowly Pulsating B and $\beta$ Cephei are $\kappa$ mechanism driven  pulsating B stars. That $\kappa$ mechanism works since a peak in the opacity due to a high number of atomic transitions from iron-group elements occurs in the area of $\log T \approx 5.3$. Theoretical results predict very few SPBs and no $\beta$ Cep to be encountered in low metallicity environments such as the Small Magellanic Cloud. However           recent variability surveys of
         B stars in the SMC reported the detection of a significant number
         of SPB and $\beta$ Cep candidates. Though the iron content plays
         a major role in the excitation of $\beta$ Cep and SPB pulsations,
         the chemical mixture representative of the SMC B stars such
         as recently derived does not leave room for a significant increase
         of the iron abundance in these stars. Whilst abundance of iron-group
         elements seems reliable, is the opacity in the iron-group
         elements  bump underestimated?
         We determine how the opacity profile
         in B-type stars should change  to excite SPB and  $\beta$ Cep
         pulsations in early-type stars of the SMC.
}

%Since the early nineties it is accepted that the excitation mechanism
 %        of B-type pulsators on the main sequence is due to the opacity
    %     peak in the Iron-group elements at $T\approx 200,000$ K.  While
       %  theoretical non-adiabatic computations predict no $\beta$ Cep
        % pulsators and  few SPBs for low metallicity environments such
         % as the Magellanic Clouds (MCs),

\maketitle

\section{Introduction}

%The excitation of modes %by $\kappa$ mechanism 
%in main-sequence B-stars has long been questioning our knowledge of these stars. The physical parameters employed, as for e.g. stellar opacities, were subject of particular attention. Only when new opacity calculations were released by the OPAL  project (Iglesias \& Rogers 1991) could the mechanism of excitation of MS B-stars be identified. These new values for stellar opacities showed a 300\% increase at $\log T\approx5.3$ , a peak in the opacity due to Iron-group elements. This peak was determined to be responsible for excitation of $\beta$ Cep (Moskalik \& Dziembowski 1992) and SPB (Dziembowski \& Moskalik \& Pamyatnykh 1993) by $\kappa$ mechanism.
The Small Magellanic Cloud is a dwarf galaxy satellite of our galaxy, that presents very low metallicity: $Z\approx0.001-0.004$ (Buchler 2008). 
Recent theoretical study (Miglio et al. 2007a) employing the updated solar chemical mixture AGS05 (Asplund, Grevesse \& Sauval 2005) find that at Z= 0.005, no excitation of $\beta$ Cep modes and a few excited SPB modes in B stars are expected.
Therefore SPBs are expected to be rare while $\beta$ Cep are thought to be absent in the SMC. However, recent detections of B-type pulsator candidates in the SMC (e.g. Karoff et al. 2008; Diago et al. 2008; Sarro et al. 2009) are now challenging our understanding of the main-sequence B-type pulsators.

In a previous work (Salmon et al. 2009), we tried to explain the existence of these B-type pulsators in the SMC by considering a chemical mixture representative of the SMC B stars. From a synthesis of the literature, we derived it and employed it to compute new theoretical models. Our results indicated that the minimal values of Z to obtain excitation respectively of SPB and $\beta$ Cep are both over the mean Z value of SMC B stars, even when the maximal error box value is taken into account.
%Thus we need to follow another approach to explain presence of B-type pulsators in the SMC.

 %We are now considering the possibility of an underestimation of stellar opacities. This
%We present here the new reasoning we have followed and that has already been adopted in the past under similar circumstances.
We foresee another possibility to reconcile observation and theory. The mechanism of excitation in $\beta$ Cep remained for long unknown: Simon (1982) suggested that a revision of opacity values for heavy elements by a factor 2 or 3 could explain excitation in $\beta$ Cep stars that was unexplained at that time. Several years later, new computations of opacity by OPAL group (Rogers \& Iglesias 1992) led to an increase by a factor 3 of the opacity due to iron-group elements, hence resolving B-type pulsators problem. More recently, an underestimation of current stellar opacity values has already been invoked to solve other remaining problems such as the excitation of low and high frequencies in galactic $\beta$ Cep pulsators, the discrepancy between the solar model with the updated solar mixture and helioseismic data and the modeling of the slowly-rotating $\delta$-Scuti 44 tau (for a review, see Montalb\'{a}n \& Miglio 2008).
% More recently it was shown that a local enhancement of the iron abundance could reproduce theoretically all the observed (i.e. excited) modes in the hybrid $\beta$ Cep-SPB pulsator $\nu$ Eri (Pamyatnykh \& Handler \& Dziembowski, 2004). Miglio (2007b) also studied the effect of a parametrical increase of iron abundance in region of exciation of B stars and its consequence on excitation in low metallicity case. 

These studies raise the following question: is stellar opacity still underestimated nowadays? 
In this work we estimate by which factor the opacity should be increased to excite pulsations in SMC B stars. We proceed by modifying the opacity values used in the theoretical computations.

We briefly recall our previous results in Section 1. In Section 2, we explain the way we proceed to modify the opacity in our models. We then present the results of our non-adiabatical computations in Section 3 and the consequence of an increase of the opacity on the excitation of B-type pulsators.

%We develop in this work a similar approach but actually we modify the opacity values used in the theoretical computations.
%Our way of proceeding allows us to emphasize directly the consequences of a modification in the employed fundamental physics, without modifying chemical profile in stellar models.
% ...Another point to take into account is a possible underestimation in the computation of stellar opacities. Stellar opacities have been calculated by two groups, namely the OP ones (Badnell et al. 2005) and OPAL ones (Iglesias \& Rogers 1996). In this work we modify the OP opacities, increasing them parametrically in the area of $\log T=5.3$, where occurs a peak in the opacity due to an important number of atomic and ionic transitions of the Iron.
\section{Effect of the chemical mixture}

Motivated by the apparent inconsistency between observation and theory for B-type pulsators in the SMC, we analyzed the effect of considering a chemical mixture representative of SMC B stars (Salmon et al. 2009). After a careful and critical review of the literature of SMC age-metallicity relation on the one hand and of chemical abundances determination of young objects (B stars and cool supergiants) on the other hand, we determined two main features for young SMC objects:\\
\begin{figure*} 
\begin{center}
\includegraphics[height=.25\textheight]{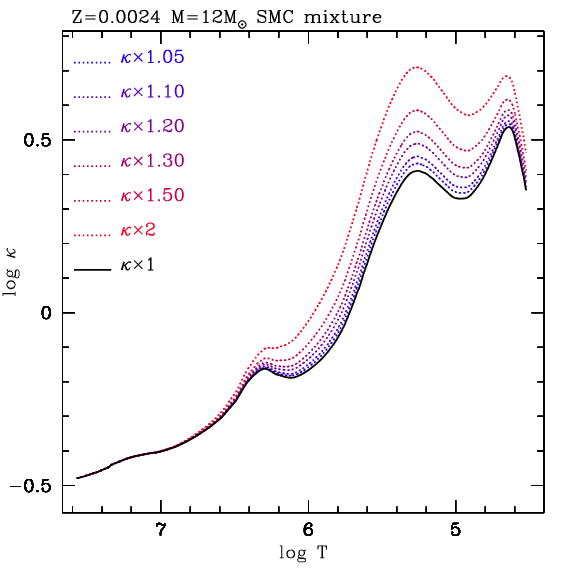}
%\hspace{2cm}
\includegraphics[height=.25\textheight]{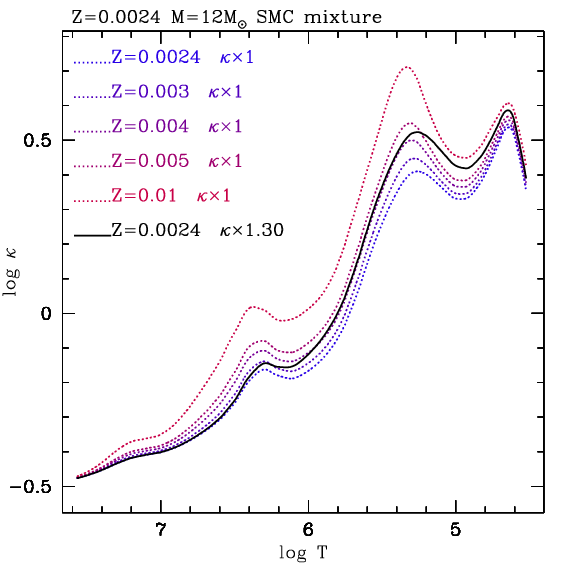}
\end{center}
\caption{Left panel : profiles of the enhanced opacities for a given structure ($\rho$ and $T$ given for a $12M_{\odot}$, Z=0.0024,Y=0.28 on ZAMS model). The greater is the $A$ factor, the wider is the opacity peak. Right panel : same as to the left, except that the 30\% increased opacity at Z$_{\textrm{{\tiny SMC}}}$ is compared to opacities at different Z, but not parametrically modified.} 
\label{fig1} 
\end{figure*}
\begin{itemize}
\item one expects [Fe/H]=-0.5$\pm$0.2 with respect to AGS05 solar chemical mixture for the youngest stars of the SMC (Carrera et al. 2008 and ref. therein) due to their age and localization: the SMC B stars benefit from the largest chemical enrichment among all the stars of the galaxy.
\item the chemical mixture representative of the SMC B stars (see Salmon et al. 2009), implying Z$_{\textrm{{\tiny SMC}}}$=0.0024$\pm$0.0006 with [Fe/H]=-0.7$\pm$0.2. The iron content we derived is in agreement with the one based on age-metallicity relations previously quoted.
\end{itemize}

We thus computed models with the stellar evolutionary code CLES (Code Li\'{e}geois d'Evolution Stellaire; Scufflaire et al. 2008), using OP opacities determined for the SMC B stars chemical mixture. The non-adiabatical analysis was performed with the MAD code (Dupret 2001). 

For models with masses ranging from 2.5 to 16 M$_{\odot}$, we did not obtain excited modes for Z$_{\textrm{{\tiny SMC}}}$.
Moreover, we did the same analysis on models of 4 and 12 M$_{\odot}$, respectively representative of SPB and $\beta$ Cep stars, for different values of Z. The results 
% These results are illustrated in the Fig. \ref{fig1}, where we represented 
% This later is expressed by :
%\begin{equation}
%\eta=\frac{W}{\int_0^R\left|\frac{dW}{dr}\right|dr}
%\end{equation}
%i.e. it represents the growth of rate of a mode.
led to the conclusion that Z should be greater than 0.004 (corresponding to [Fe/H]$\sim$-0.45) to obtain excited SPB-type modes and greater than or equal to 0.007 ([Fe/H]$\sim$-0.2) to obtain excited $\beta$ Cep-type modes. 

\section{Parametric increase of the opacity}

As we cannot explain the existence of $\beta$ Cep pulsators (SPB being marginally compatible) in the SMC employing Z$_{\textrm{{\tiny SMC}}}$ and a representative chemical mixture in our models, the following question follows: is the stellar opacity underestimated? Recent studies explored the effect of increasing opacity by enhancing the iron abundance in stellar models (see Ausseloos et al. 2004; Pamyatnykh, Handler \& Dziembowski 2004; Miglio et al. 2007b). In the same idea we proceed to an \emph{ad hoc} and local change in the opacity used in our models, in order to estimate which enhancement of the opacity could lead to excitation of SPB- and $\beta$ Cep-type modes at Z$_{\textrm{{\tiny SMC}}}$. %We could have increased abundances in the area of iron peak opacity which mimics an increase of the latter but it also changes locally chemical mixture in the model. 
Our way of proceeding allows us to emphasize directly the consequences of a modification in the employed fundamental physics.

Iron is the element mainly contributing ($\approx 67 \%$) to the peak of opacity in the layers with $T\approx 200,000$ K, responsible for excitation in B stars. Thus we decided to consider an increase of the opacity which would be due to an underestimation of the iron contribution.
We proceed by modifying the opacity (OP one, calculated for the SMC B stars mixture) as follows:
\begin{equation}
\kappa'(T)=\kappa(T) \times \left\{1+A\ \exp \left[ -\left(\frac{T-T_{\textrm{{\tiny Fe}}}}{2\sigma_{\textrm{{\tiny Fe}}}}\right)^2\right]\right\}
\end{equation}
where $\kappa$ is the Rosseland mean opacity and $T_{Fe}\ \textrm{and}\ \sigma_{Fe}$ are respectively equal to $5.30$ and $0.5$. These values have been chosen to match accurately the contribution of iron to the opacity in the area of $T\approx200,000 K$.
% where runs the mechanism of excitation. 
The factor of increase, $A$, has been set to different values, namely 0.05, 0.10, 0.20, 0.30, 0.50 and 1. The modified opacities are represented in Fig. \ref{fig1} (left panel).  %the profile of opacity increased by 30 \% at Z$_{\textrm{{\tiny SMC}}}$ if compared to the non modified profiles at different metallicities in Fig. 2 (right part). 
The modified iron peak opacity value is equivalent to a case between $Z$=0.004 and $Z$=0.005 for unchanged opacities, as illustrated in Fig. 1 (right panel). However the modified profile has a different slope in the iron bump area than in the case of unchanged profiles, this slope impacting on the damping of modes. The excitation depends not only on the value of opacity near $\log T$=5.3 but also on its change with $T$.  %On the right part, the opacity increased by a factor 1.30 evaluated for a given structure ($\rho,T$ fixed from a model at $Z_{\textrm{\tiny SMC}}=0.0024$ and $Y=0.28$) is compared to the non-increased opacities for the same structure but assuming different values of Z when evaluated. The value of the Iron peak for the new opacity is slightly higher than in the case $Z=0.004$. Nonetheless the modified opacity has a Iron peak wider, presenting a sharper derivative on the left wing (the hotter one) and a smoother derivative on the right wing (the colder one).

\begin{figure*} 
\begin{center}
\includegraphics[height=.25\textheight]{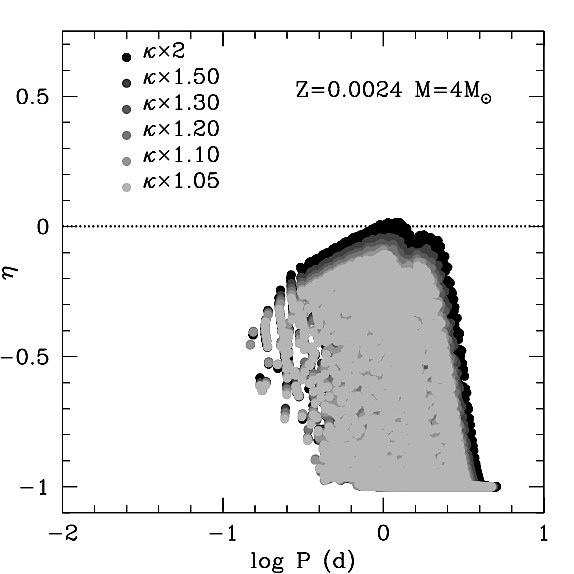}\includegraphics[height=.25\textheight]{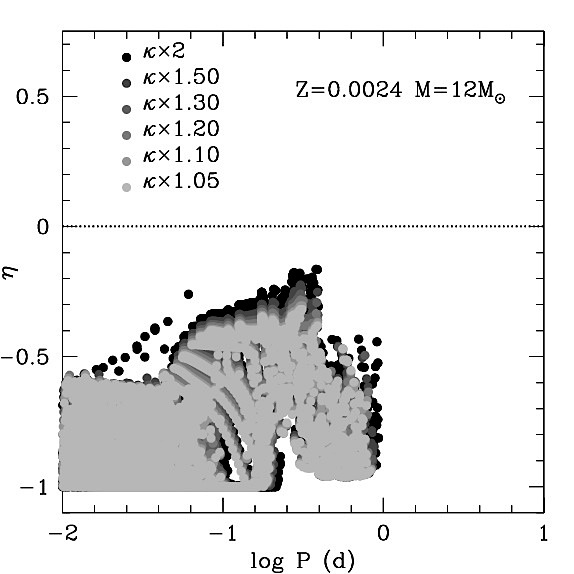}  
\end{center}
\caption{The $\eta$ parameter of modes (reported to the logarithm of their period) for a typical SPB (resp. $\beta$ Cep) model of the SMC in the left (resp. right) panel. The $\eta$ parameter indicates the rate of growth of a mode. A positive value (resp. negative) indicates that the mode is excited (resp. damped). 
} 
\label{fig2} 
\end{figure*}

\section{Stability analysis for models with modified opacity}

\begin{figure} 
\includegraphics[height=.32\textheight]{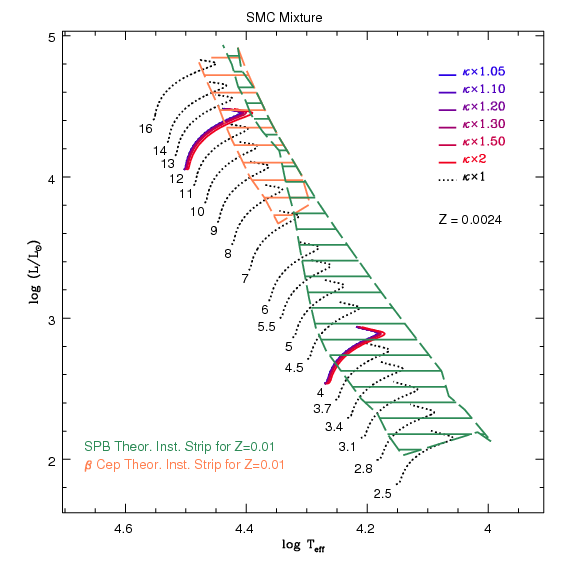}
\caption{Evolutionary tracks for non-modified and modified opacity models in the HR diagram. Indicative instability strips of SPB and $\beta$ Cep-type pulsations are obtained from models computed with $Z$=0.01 and $\kappa \times 1$.} 
\label{fig3} 
\end{figure}

We calculate models of 4 and 12 M$_{\odot}$, using the opacities which we modified for the different factors of increase $A$ and $Z$=0.0024, 0.003, 0.004, 0.005 and $0.01$. In Fig. \ref{fig3}, we present in a HR diagram the main sequence evolutionary tracks of models for masses between 2.5 and 16 M$_{\odot}$, adopting unchanged opacity. On the same figure are represented the 4 and 12 M$_{\odot}$ models at Z$_{\textrm{{\tiny SMC}}}$ computed with the different increased opacities. For both masses, the position of main sequence tracks in the HR diagram is not considerably affected.
% is shift the more the opacity is increased, the more the track is shifted to the right i.e. to colder effective temperatures.\\

%fe is the element mainly contributing ($\approx 67 \%$) to that peak of opacity. Thus fe is only taken into account here to modify the total opacity under the assumption that its contribution is underestimated.

Results of the non-adiabatic calculations are shown for the 4 and 12 $M_{\odot}$ models at Z$_{\textrm{{\tiny SMC}}}$ in the left and right panels of Fig. \ref{fig2} respectively. It appears that SPB modes are excited for an opacity multiplied by 2 ($A$=1), while no $\beta$ Cep modes are excited for the factors of increase $A$ we used. In Fig. 4 (left panel), one remarks that excited g modes are found at the end of the evolution on the main sequence for the SPB model in which the iron peak is increased by a factor 2.
%No low-order p modes are excited for this model representative of a SPB. For the $12M_{\odot}$, neither p nor g modes are excited for Z$_{\textrm{{\tiny SMC}}}$ whatever is the augmentation of the opacity.
Excited frequencies of the $4M_{\odot}$ model at Z$_{\textrm{{\tiny SMC}}}$ and for an increase $A$=1 are compared to the observational frequencies of the detected SMC SPBs in Fig. 4 (right panel).

\begin{table}
% \centering%%%
\caption{ Value of the increasing factor $A$ necessary to obtain excitation of p or g modes is reported for the different combinations of model parameters ($Z$ and $M$).}
\label{table1}
\begin{tabular}{lllll}\hline
& M=4 M$_{\odot}$ & &M=12 M$_{\odot}$ & \\
& g modes & p modes & g modes & p modes \\
\hline
Z=0.0024 & 1 & / & / & /\\

Z=0.003 & 0.50 & / & / & / \\
Z=0.004 & 0.10 & / & / & / \\

Z=0.005 & 0 & / & 1 & 1 \\ 

Z=0.01 & 0 & / & 0 & 0\\
\hline
\end{tabular}
\end{table}

\begin{figure*} 
\hspace{0.4cm}
\includegraphics[height=.3\textheight]{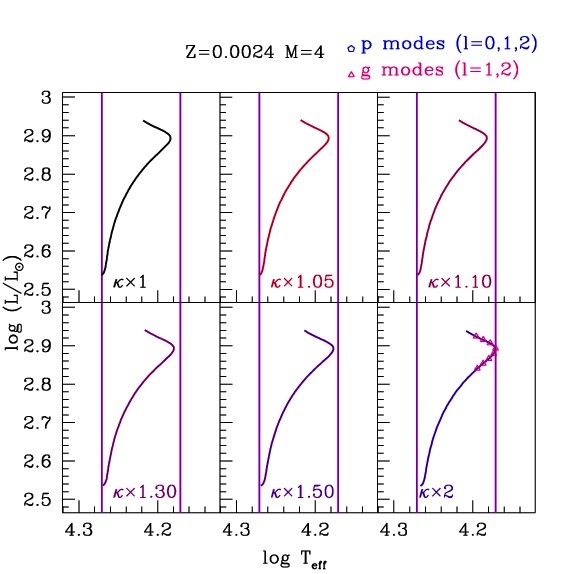}
\hspace{0.5cm}
\includegraphics[height=.275\textheight]{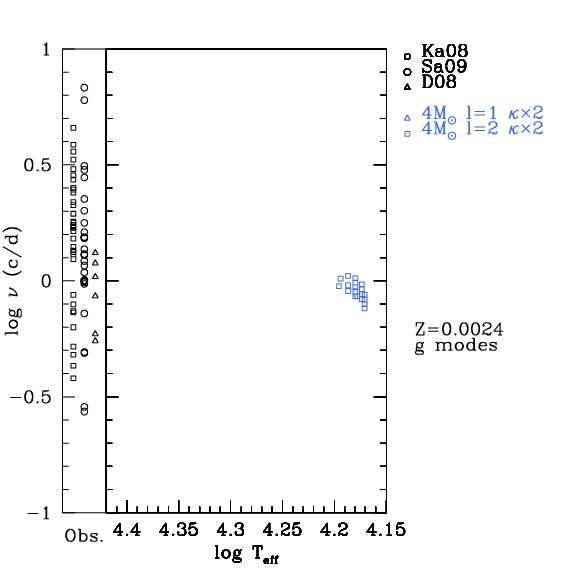}
\caption{Left panel : Evolutionary tracks for a 4 M$_{\odot}$ model computed for several increases in the opacity Iron peak. Models with theoretically excited p or g modes are marked by a pentagon or a triangle. Right panel : on the left part of the figure is represented the logarithm of the frequency (in c/d) for SPB modes observed in the SMC (Diago et al. 08; Karoff et al. 08 and Sarro et al. ). On the right part is reported the logarithm of the frequency to the $\log T_{eff}$ of the model in which the mode is found excited.} 
\label{fig4} 
\end{figure*}

Finally, we report in Table \ref {table1} the minimal factor of increase $A$ that excites p or g mode in the 4 or 12 M$_{\odot}$ models for the metallicities we considered. We find excited g modes in the case $Z$=0.0024, $M$=4 M$_{\odot}$ and $A$=1. When considering $Z$=0.003 (upper limit within the 1-$\sigma$ error box of the Z of the SMC B stars), g modes are excited in the 4 M$_{\odot}$ model for $A$=0.50. 
A slight increase (10\%) of the opacity at $Z$=0.004 excites g modes in the 4 $M_{\odot}$ model. %This was expected when reminding the Fig. \ref{fig1}, where this mass is at the very limit to obtain SPB-type modes.
Excited p modes are only found in the $\beta$ Cep model (12 M$_{\odot}$), appearing at $Z$=0.005 for a factor 2 ($A$=1) of increase in the opacity. For the same conditions, excited high-order g modes are also obtained.

Under the gaussian profile assumption that we have made to modify the opacity profile, one remarks that increasing the opacity in the iron peak does not change the intrinsic nature of modes excited in our models: SPB and $\beta$ Cep representative models remain of the same nature.  In the 12 M$_{\odot}$ model at $Z$=0.005 and $A$=1 as excited low order p modes appear, so do high order g modes. From an observational point of view, considering opacities as underestimated by more than 50\%, SPB modes should be detected in SMC B stars with Z$_{\textrm{{\tiny SMC}}}$=0.0024 ([Fe/H]=-0.7) while $\beta$ Cep modes should be observed in the stars with $Z$=0.005 ([Fe/H]=-0.35), i.e. a metallicity within 5-$\sigma$ error box on Z$_{\textrm{{\tiny SMC}}}$.  

%We do not distinguish if this soon hybrid character of the $\beta$ Cep model is a result of the way the opacity is modified, or a characteristic property (as for models at $Z=0.01$ with non-modified opacity).

\section{Conclusion}

We have investigated the effect on the excitation of modes in SMC B stars of an underestimation of stellar opacity values in the peak due to iron. We proceed by multiplying the opacity used in our model computations by a gaussian profile centered on the temperature where contribution of iron to opacity is maximum, i.e. $\log T$=5.3. The results indicate that at the metallicity of the SMC B stars, Z$_{\textrm{{\tiny SMC}}}$=0.0024, SPB modes could be excited for an augmentation greater than 50\% of the opacity. No $\beta$ Cep modes are excited at Z$_{\textrm{{\tiny SMC}}}$ for an increase from 5\% to 100\% of the opacity. Nevertheless, $\beta$ Cep modes are found excited for an increase greater than 50\% but at $Z$=0.005, that is 5-$\sigma_{{\tiny Z}}$ ($\sigma_{{\tiny Z}}$=0.0006) away from the metallicity we derived for SMC B stars.
These values of increase must be taken with care since they are obtained for opacity modified according to an \emph{ad hoc} defined mathematical function that determines the slope of the opacity profile. The impact of the choice of mathematical function  and its consequence on the pulsation driving efficiency will be refined in the future.
%In this work we determine the factor of increase, $A$, necessary to excite high-order g modes at Z$_{\textrm{{\tiny SMC}}}$ in SPB model: factor A should be strictly higher than 1.5. For the same metallicity we do not excite either p or g modes in the $\beta$ Cep model. 
%Within the error on the estimation of the Z$_{\textrm{{\tiny SMC}}}$, it appears that a factor $A$=1.50 could be enough to obtain excited g modes in SPB model.

%Low-order p modes are found excited in the $\beta$ Cep model for an increase $A$=2 when taking $Z$=0.005. In the same time high-order g modes are found to be excited in this model. %To clearly identify the origin of this hybrid character, a careful attention on the internal structure of our models will be bring in the near future.

The stellar depth where stands the opacity peak also plays a role both on the damping of modes and on the excited frequencies spectrum in B stars. Other elements of the iron-group (Ni, Cr and Mn), contribute to that bump in the opacity at slightly different temperatures affecting position of the opacity peak. If we suppose an underestimation of the iron opacity, it would be reasonable to consider the same for the other iron-group elements opacity. Thus we will take them into account in the next step of our theoretical modeling of SMC B stars.

\end{document}